\documentclass[10pt,aps,prl,superscriptaddress,
twocolumn,showpacs,floatfix,notitlepage]{revtex4-2}
\usepackage{amsmath}
\usepackage{amsfonts}
\usepackage{amssymb}
\usepackage{hyperref}
\usepackage{graphicx}
\usepackage{color}
\usepackage[mathscr]{euscript}
\usepackage{bbm}
\usepackage{braket}
\usepackage{amsmath}
\usepackage{amssymb}
\usepackage{wrapfig}
\usepackage{tabularx}
\usepackage[utf8]{inputenc}
\usepackage{booktabs}
\usepackage[table]{xcolor} 
\usepackage{siunitx}

\usepackage{natbib}
\usepackage{rotating}
\usepackage[utf8]{inputenc}
\usepackage[T1]{fontenc}
\usepackage{soul}

\hypersetup{colorlinks=true,linkcolor=magenta
  ,citecolor=magenta, urlcolor=magenta}

\begin{document} 
\keywords{Slow light, Group velocity dispersion, Semiconductor excitons, Dielectric superlattice, Perovskite waveguides}
\title{
Strongly Nonlinear Slow Light Polaritons in Subwavelength Modulated Waveguides}
\author{Amir Rahmani}
\affiliation{%
	Institute of Physics, Polish Academy of Sciences, Al. Lotnik\'{o}w 32/46, 02-668 Warsaw, Poland}%
\affiliation{%
	Center for Quantum-Enabled Computing, Center for Theoretical Physics, Polish Academy of Sciences, Al. Lotnik\'{o}w 32/46, 02-668 Warsaw, Poland}%

\author{Maciej Dems}
\affiliation{%
	Institute of Physics, Lodz University of Technology, ul. Wólczańska 217/221, 90-003 Łódź, Poland}%

\author{Michał Matuszewski}
\affiliation{%
	Institute of Physics, Polish Academy of Sciences, Al. Lotnik\'{o}w 32/46, 02-668 Warsaw, Poland}%
\affiliation{%
	Center for Quantum-Enabled Computing, Center for Theoretical Physics, Polish Academy of Sciences, Al. Lotnik\'{o}w 32/46, 02-668 Warsaw, Poland}%

\begin{abstract}  
Slow light is a regime of reduced group velocity, resulting in increased photon density in optical pulses and enhanced nonlinear effects.
Here, we propose the realization of slow light in the regime of strong light-matter interaction between waveguide photons and semiconductor excitons. 
We  design a dielectric superlattice structure with a nearly-flat band characterized by low group velocity and group velocity dispersion, both required for enhancing nonlinear effects with ultrashort pulses. Furthermore, by applying this general framework to a perovskite-based structure, we demonstrate an enhancement of the single-particle phase shift by a factor of more than 20, representing a significant step toward the few-photon quantum regime. Our results provide a blueprint for accessible strong interactions in solid-state integrated optics.
\end{abstract}

\date{\today}

\maketitle
\emph{Introduction.}\label{sec:intro}
Nonlinear optical effects are pivotal for a wide range of modern photonic applications~\cite{Maring2024,Keni25,chen13}. They are a crucial for optical computing and optical neural networks~\cite{shastri2021photonics,opala2023harnessing}. Of particular interest is the regime of strong single-photon interactions, where the optical nonlinearity is sufficient for a single photon to change the state of another one, for instance, by imprinting a phase shift~\cite{PhysRevLett.62.2124}. This capability is essential for realizing deterministic optical quantum gates and generating entangled photon states \cite{Chang2014,Yanagimoto24}. Due to the vanishingly small interaction between photons in free space, matter-mediated schemes are used to induce effective optical nonlinearities~\cite{Kala2025}. Particularly strong nonlinearities  result from coupling photonic modes to atomic~\cite{Volz2014,Tiarks16} or solid-state emitters~\cite{Ilya08,Delteil2019,Staunstrup2024,benimetskiy2025all,Makhonin2024,Kuriakose2022}, or using mechanisms such as electromagnetically induced transparency~\cite{Feizpour2015}.

Slow light provides an effective route to enhance light–matter interactions~\cite{Krauss2008,Benis:22,Schmidt:96,Krauss2007}. It can be realized by dispersion engineering~\cite{ZHANG2022128721,Changhao19} in dielectric structures, by material resonances~\cite{lorentz1909theory,Hau1999} and by topological phases~\cite{Kumar2024}. Many nonlinear optical devices rely on the generation and control of an optical phase shift $\Delta \phi$~\cite{Baba2008,Khurgin2023}. In the absence of group velocity dispersion, optical mode in the slow-light regime experiences phase shift enhancement over a fixed propagation length $L$ that scales inversely with the group velocity $v_g$, giving $\Delta \phi \propto \frac{L}{v_g}$. In addition, the optical mode is spatially compressed in the direction of propagation, which results in increase of local field intensity. This  enhances the nonlinear response even further, providing an additional $v_g^{-1}$ factor. In result, accumulated  phase shift scales as $\Delta \phi \propto \frac{L}{v_g^2}$, implying that slow light propagation can lead to much larger nonlinear phase shifts. 

Nonlinear optical effects can be further enhanced in the strong light–matter coupling regime~\cite{Ge21,Wang2021}. This regime is reached when the coherent exchange of energy between an optical mode and a material excitation occurs faster than the relevant decoherence processes, such as dissipation. Under these conditions, the photon and the excitation hybridize to form dressed modes known as polaritons~\cite{BasovAsenjo}. Owing to their mixed photonic and material character, polaritons inherit strong optical nonlinearities from the matter component. Moreover, dimensional confinement can reduce the effective mode volume, and a high-quality factor extends the interaction time, further enhancing the nonlinear optical response. To date, nonlinear phase shifts per particle $\Delta \phi_{\text{single}}$ in the strong coupling regime have been measured   experimentally with zero-dimensional open cavity exciton-polaritons ($\Delta \phi_{\text{single}}\approx0.08\pi$ rad)~\cite{benimetskiy2025all}, with molecules placed in a microcavity ($\Delta \phi_{\text{single}}\approx0.37\pi$ rad) \cite{Wang2019}, cloud of atoms in a cavity ($\Delta  \phi_{\text{single}}\approx\pi$ rad) \cite{PhysRevX.12.021034}, quantum dot in a photonic crystal cavity ($\Delta \phi_{\text{single}}\approx0.25\pi$ rad) \cite{Ilya08};   however, these have not yet led to a scalable integrated photonics platform. 

Here, we explore the strong-coupling regime with structurally shaped slow light. We propose a strategy to realize exciton-polariton slow light by engineering the dispersion of a subwavelength dielectric waveguide. The superlattice structure has a fishbone geometry with embedded defects, giving rise to a near-flat photonic band. When coupled to excitonic emitters, this configuration enables the formation of slow polaritonic modes in the strong light–matter coupling regime. We predict a significant enhancement of the nonlinear phase shift by a factor exceeding 20$\times$ compared to a regular waveguide, using a simple design compatible with experimentally accessible perovskite structures~\cite{Kedziora2024}. While these materials generally exhibit weak third-order nonlinearity~\cite{PhysRevMaterials403}, our results show that integrating them with slow-light modes can significantly enhance their effective nonlinear response. The proposed geometry is scalable and provides a promising route for integrating nonlinear elements on semiconductor photonic chips at room temperature. To the best of
our knowledge, we present the first proposal for a giant phase shift in the strong coupling regime that uses
structural dispersion engineering. 

It is worth mentioning that bound states in the continuum (BICs)~\cite{Yu2023} can also lead to enhanced nonlinearity~\cite{Fang2025}. Importantly, while in our case slow light modes are conventional guided modes with a finite bandwidth, residing below the light cone, BIC modes lie inside the radiation continuum but remain decoupled from it thanks to a particular symmetry. As a result, BIC modes are not protected by total internal reflection and can be more sensitive to to disorder-induced symmetry breaking.

\begin{figure}
    \centering
\includegraphics[width=0.99\linewidth]{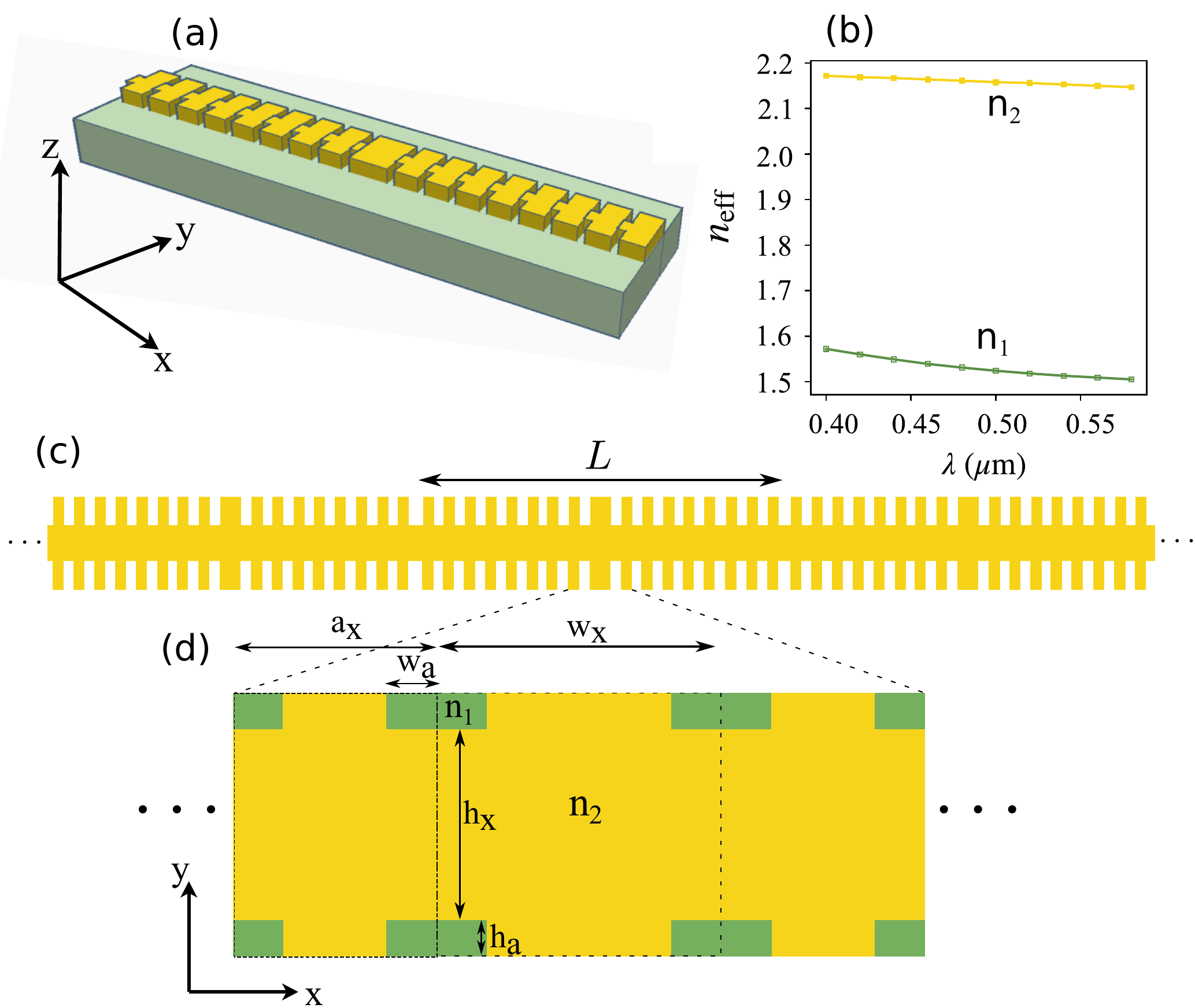}
\caption{Schematic of the designed slow polariton perovskite waveguide.
(a) In a strip single-channel waveguide, the guiding layer is corrugated and includes one defect per supercell, whose periodic repetition forms a superlattice. (b) Effective refractive indices of the first TE fundamental mode for the geometry shown in panel (a) in etched ($n_1$) and raised ($n_2$) sections. (c) In the top view (xy plane), the geometry of superlattice reduces to a 2D problem. $L$ shows the size of the unit cell (supercell) in the superlattice. (d) Close-up view of the defect region and the two unit cells. The defect has the width $w_x$, and the period of the corrugation is $a_x$.\label{fig1:fih1mixing}}
\end{figure}

\emph{Optical modes and slow light.} We investigate the sample shown schematically in Fig.~\ref{fig1:fih1mixing}, which is a corrugated raised waveguide on a substrate. This structure consists of a periodic repetition of a basic corrugated unit cell with period $a_x$, together with a defect of size $w_x$. It therefore forms a superlattice with unit cell length $L=N_ga_x+w_x$
, where $N_g$ is the number of basic unit cells between successive defects. The resulting waveguide can be viewed as a coupled-resonator optical waveguide~\cite{Yariv99,Changhao19}. This structure can be implemented e.g.~using lead-based perovskite on glass~\cite{Kedziora2024}. Our initial objective is to determine the eigenmodes and energy-momentum dispersion for this structure. In the case of an isotropic material, this requires solving the eigenvalue problem for the full 3D wave equation: $\nabla\times\nabla\times \mathbf{E}(x,y,z) + k_0^2 n^2(x,y,z) \mathbf{E}(x,y,z) = 0$, where $\mathbf{E}$ represents the electric field vector and $n(x,y,z)$ is the refractive index profile of the geometry. To reduce computational complexity, we use the effective index method \cite{Chiang86,Hammer2009}, which approximates the 3D problem as an effective 2D system \footnote{This approximation is valid when the higher-order modes of the waveguide are not excited significantly, or if their energy separation is larger than other energy scales.}. In 2D $x$-$y$ plane we have $\left( \frac{\partial^2}{\partial x^2} + \frac{\partial^2}{\partial y^2} \right) F(x,y) +  [k_0^2 n^2_{\text{eff}}(x,y)-\beta^2] F(x,y) =0$, where $n_{\text{eff}}$ is given for each $(x_i,y_i)$ pair by the solution  of $\frac{d^2 F(x_i,y_i,z)}{dz^2} + [k_0^2 n^2(x_i, y_i, z)-k_0^2 n_{\text{eff}}] F(x_i,y_i,z) = 0$ where $F(x_i,y_i,z)$ represents a transverse eigenmode, for instance a TE or TM mode. To calculate the eigenmodes and eigenvalues we use the
plane wave admittance transfer method \cite{Dems05} implemented
in the PLaSK software~\cite{plask}. The results of eigenmode analysis are presented in Fig. \ref{fig1:fih1mixing}(b), where we plot two refractive indices of the fundamental modes: $n_2$ (yellow curve) for the central region and $n_1$ (green curve) for the etched sections, see the schematic in panel (c). 
\begin{figure*}
    \centering
\includegraphics[width=0.9\linewidth]{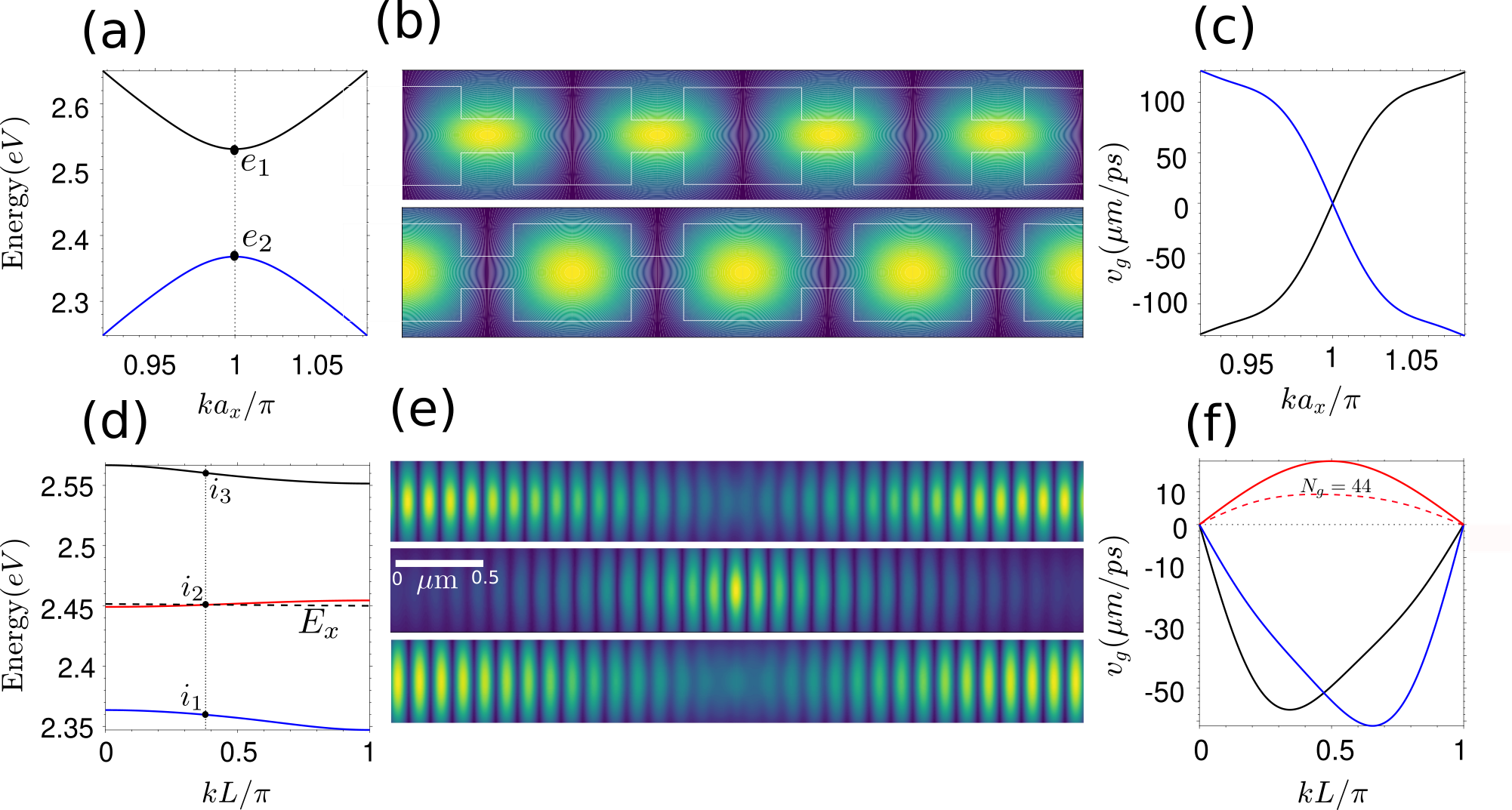}
    \caption{ Upper row: dispersion and modes of the periodic corrugated structure without defects.  (a) Periodic modulation introduces a bandgap with upper and lower TE bands  separated by a gap at the band edge ($ka_x=\pi$). (b) Electric field intensity profiles of TE modes corresponding to lower and upper modes at the band edge (points $e_1$ and $e_2$, respectively). (c) Group velocities of the two bands, vanishing at the band edge. Lower row: dispersion and modal properties of the corrugated structure with defects (superlattice). (d) Photonic dispersion of the upper (black), gap (red), and lower (blue) bands for a supercell of length $L$. The middle band is nearly flat. The horizontal dashed line shows the exciton energy. (b) Examples of electric field profiles corresponding to each band (points $i_1$, $i_2$ and $i_3$) in panel (a). In the flat band, the field is localized at the defect (centered). (c) Group velocities in each band. The solid curves show group velocities corresponding to the band mode in (d) with $N_g=32$. The dashed curve show the group velocity for $N_g=44$.  Parameters are $a_x=0.13\,\mu $m, $h_a=0.08\,\mu $m, $h_x=0.3-2h_a\,\mu $m, $w_a=0.02\,\mu $m, $w_x=0.07\, \mu$m. \label{fig2:fihsdcssdd1mixdscing}}
\end{figure*}

Using the effective index method, we now study the dispersion and modal properties of the 2D corrugated fish-bone structure. The results are presented in Fig.~\ref{fig2:fihsdcssdd1mixdscing}. We begin the analysis with an ideally periodic structure without defects ($w_x=a_x$), see panels (a)-(c). Panel (a) shows that the corrugation along the $x$ direction opens a photonic band gap at the edge of the Brillouin zone (BZ)  $k=\pm\pi/a_x$ due to Bragg scattering. The corresponding TE modes at the band edge are shown in panel (b), which reveals two modes with distinct spatial localization: the lower band mode localizes in the high-index rib regions, whereas the upper-band mode resides in the low-index gaps. This implies that group velocity $v_g$ vanishes at the Brillouin zone edge and increases rapidly away from it, see panel (c). Moreover, the periodic geometry exhibits relatively large group-velocity dispersion which leads to fast pulse spread, decay of the amplitude, and limits nonlinearity for long propagation distance. To mitigate the large group velocity and group-velocity dispersion, we introduce structural defects in a superlattice by modifying the width of one of the cells to 
$w_x\neq a_x$, thereby forming a supercell of length $L$. As shown in the supercell dispersion of panel (d), a  defect-induced band (red curve) emerges within the fundamental photonic band gap. The gap mode is well described by the known dispersion of a coupled-resonator optical waveguide, $E_g=\hbar \omega_g\left(1+\kappa \cos(kL)\right),$ where $\hbar \omega_g=2.4527~\mathrm{eV}$ and $\kappa=-0.0011$ is the coupling constant~\cite{Yariv99}. The field profiles in panel (e) show the extended modes of the lower and upper bands and the localized gap mode. As expected for a state lying inside the photonic band gap, the defect mode strongly confines the optical field at the defect site. This localization produces a slow light regime, as shown in panel (f), where the group velocity is strongly reduced over a broad frequency range. Namely, the defect band is nearly flat (compared to the upper and lower band), resulting in a much smaller group velocity dispersion than in the pristine, preiodic lattice. It is worth noting that increasing the corrugation within the supercell improves the band flatness. An example is shown in panel (f), where the gap mode for $N_g = 44$ is marked by dashed lines.

%

\emph{Giant nonlinear phase shift.} 
Following our analysis of the slow-light regime in the corrugated waveguide structure with defects, we now turn to determining the slow-light enhancement of the nonlinear phase shift under strong light–matter coupling. Generally, the magnitude of the nonlinear phase shift is set by the interplay between the intrinsic nonlinear susceptibility of the medium and the effective interaction time between light and matter~\cite{Khurgin2023}. The intrinsic contribution is governed by the electronic structure and thus imposes a fundamental limit on the attainable phase shift. We consider two kinds of nonlinearities; the optical Kerr-like effect~\cite{89946}, arising from virtual transitions, and the absorption saturation and phase-space filling~\cite{klingshirn2012semiconductor} relying on carrier excitations.

Unlike these intrinsic material constraints, the interaction time constitutes a structural parameter that can be tailored. Examples are  slow-light regime, metamaterials~\cite{Litchinitser01012018,Fang2025} and exceptional points~\cite{Pick17,Huang22,Wang2024}. That said, for the material platform considered here, we focus on Wannier excitons in a three-dimensional perovskite structure strongly coupled to the slow light mode. 
We model the strong coupling of slow light to excitons using the system of equations
\begin{subequations}\label{eq:coupled92387eh}
\begin{align}
    i\hbar \partial_t \psi_c &= \hbar \hat{\omega}_c \psi_c + \hbar \Omega \psi_x - i\hbar \gamma_c \psi_c\,,  \\
    i\hbar \partial_t \psi_x &= \hbar\omega_x\psi_x + g_x |\psi_x|^2\psi_x + \hbar \Omega \psi_c - i\hbar \gamma_x \psi_x\,.
\end{align}
\end{subequations}
Here, we introduce the slow light photonic field $\psi_c$ governed by the dispersion operator: $\hat{\omega}_c =\sum_n \frac{(-1)^n(i\partial_x+k_{ex})^n}{n!}\frac{d^n\omega_c(k)}{dk^n}|_{k=k_{ex}}$ with $k_{ex}$ stands for the wavevector at which the photon dispersion crosses the exciton resonance (see panel (d) in Fig.~\ref{fig2:fihsdcssdd1mixdscing}) and $\omega_c$ is the slow light band frequency. The exciton field $\psi_x$ is characterized by a free energy $\omega_x$ and is subject to two distinct nonlinear mechanisms. The first arises from exciton-exciton interactions with interaction coefficient $g_x$. The second originates from phase-space filling, which saturates the light-matter coupling and, to the lowest order, is given by $\hbar \Omega = \hbar \Omega_0 - \beta_x |\psi_x|^2$~\cite{PhysRevB.52.7810,PhysRevB.92.235305}, where $\Omega_0$ denotes the vacuum Rabi splitting and the saturation density can be estimated as $n_{\text{sat}} \approx 0.01/ a_B^{3}$. This estimate is valid at low temperatures, while at higher temperatures the saturation coefficient may differ~\cite{klingshirn2012semiconductor}. We define the saturation parameter as $\beta_x = \hbar \Omega_0 / n_{\text{sat}}$ \cite{PhysRevB.92.235305}. In 3D perovskite structures, the interaction constant can be estimated as $g^{3D}_{x} \approx 13.6 E_B a_B^3$~\cite{keldysh1968collective}. For typical values of $a_B \approx 5 \text{ nm}$ and $E_B \approx 40 \text{ meV}$, we obtain $g^{3D}_x \approx 0.07~\mu \text{eV} \mu \text{m}^3$. Given $\hbar \Omega_0 \approx 150 \text{ meV}$, we find $\beta_x^{3D} \approx 2~\mu \text{eV} \mu \text{m}^3$, indicating that phase-space filling is almost two orders of magnitude stronger than contact interaction. In 1D model, these constants are replaced by effective coefficients rescaled by transverse mode widths, $g_x = g^{3D}_{x}/(w_y w_z)$ and $\beta_x = \beta^{3D}_{x}/(w_y w_z)$. We also assume that photon and exciton losses are quantified by decay rates $\gamma_c$ and $\gamma_x$, respectively. Modes in waveguide geometry may suffer mainly from  radiation loss, particularly for our gap mode, which is well localized in real space and therefore broadened in momentum space, leading to intrinsic decay~\cite{1017597,Povi06}. Since radiation in the $x$-$y$ plane is taken into account by the eigenmode solver, the flat band mode has a larger decay rate (imaginary part of the energy) compared to the other modes, which is on the order of a few tenths of a meV. 

\begin{figure}
    \centering
\includegraphics[width=1\linewidth]{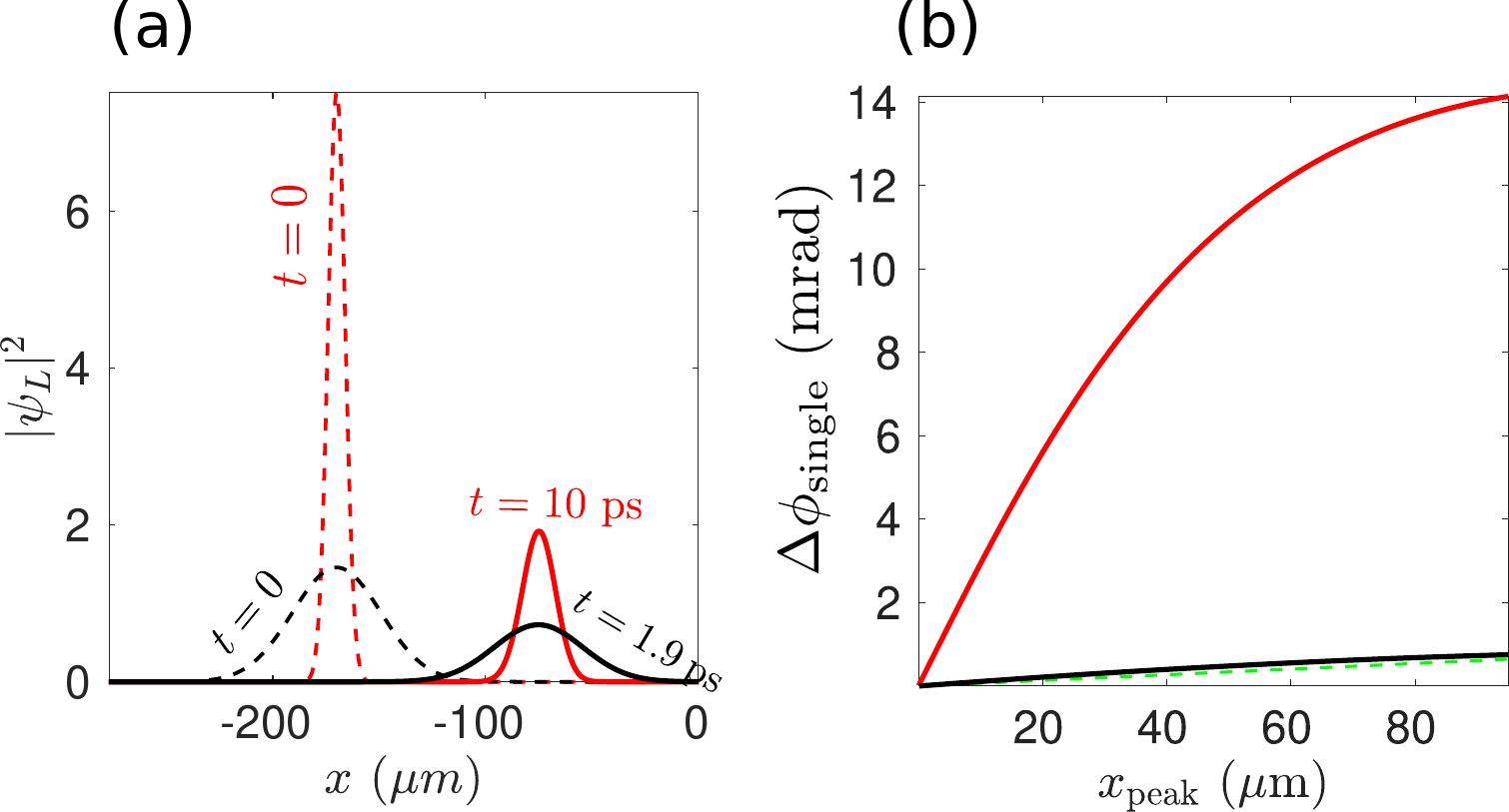}
    \caption{(a) Snapshots of eigenmode wavepacket intensity in the slow-light and fast-light regimes. Initially, there is a Gaussian wavepacket (shown in red and black dashed lines for the fast and slow light, respectively) traveling along the x-axis. The red curve shows the wavepacket intensity after 10 ps of evolution, and the black curve shows the case of fast light. (b) Average nonlinear phase shift shown in red for the gap band. The black line shows the nonlinear phase shift accumulated in the fast-light regime. The green dashed curve shows the nonlinear phase shift for a strip-raised waveguide without any corrugation. \label{fig3:fihsdsdcsdcsd1mixdscingdfv}}
\end{figure}

Our primary quantity of interest is the nonlinear phase shift, defined as $\Delta\phi=\langle \phi\rangle_{\text{lin}}-\langle \phi\rangle_{\text{non}}$ with $\langle \phi\rangle_{\text{lin}}$ and $\langle \phi\rangle_{\text{non}}$ are the linear and nonlinear intensity-weighted average phase shift, respectively, defined as $\langle \phi(t)\rangle=\frac{\int |\psi(x,t)|^2 \phi(x,t)dx}{\int |\psi(x,t)|^2 dx}$, where $\phi(x,t)$ is the unwrapped local phase of the complex field, i.e.~$\psi(x,t)=A(x,t){\rm e}^{ i\phi(x,t)}$. We assume that the field  consists of lower polaritons  ($\psi_L$), setting the initial conditions for Eqs.~(\ref{eq:coupled92387eh}) as $\psi_C(t=0,x)=A\exp(-x^2/w^2)$ and $\psi_X(t=0,x)=Ae^{i\pi}\psi_C(t=0,x)$, reconstructing the $\pi$ phase difference between exciton and photon components in lower polaritons. Under this approximation, the relative phase between $\psi_C$ and $\psi_X$ is preserved during propagation. Initial pulse profiles are shown in Fig. \ref{fig3:fihsdsdcsdcsd1mixdscingdfv}(a) (dashed curves). In the slow-light regime, pulses are spatially compressed and intensity is enhanced, whereas pulses in the normal regime are broad with smaller amplitudes. To see this, in a non-dispersive medium one may assume a Gaussian envelope $\psi_0 = A_0 \exp[-(t - x/v_p)^2 / 2\tau_0^2]$ with duration $\tau_0$, where $v_p = c/n$ is the phase velocity. Defining the slowdown factor $S = v_p/v_g$, upon entering the dispersive medium the spatial profile becomes $\psi = A \exp[-S^2 x^2 / 2w_0^2]$, where $A = A_0 \sqrt{S}$ and $w_0 = \tau_0 v_p$. Here, we assume $\tau_0=350$~ft. Figure \ref{fig3:fihsdsdcsdcsd1mixdscingdfv}(a) also compares the final pulse profiles for two distinct regimes. In the slow-light regime (red curve), corresponding to the near-flat band within the gap, the profile is shown at $t = 10$ps. In contrast, for the fast-light regime (lower band), the pulse reaches the structure's boundary at $t \approx 1.9$ps (black curve). Although the pulse amplitude decays in both cases, the attenuation is markedly reduced in the slow light regime. As the pulse traverses several supercell lengths, it accumulates a nonlinear phase. The temporal evolution (scaled with the group velocities) of the average nonlinear phase is depicted in Fig. \ref{fig3:fihsdsdcsdcsd1mixdscingdfv}(b). The phase first increases  and approaches saturation at later times, reaching approximately $14$~mrad. In the fast-light regime, the phase accumulation is minimal over propagation distances comparable to those in the slow-light case (black line, Fig. \ref{fig3:fihsdsdcsdcsd1mixdscingdfv}(b)). Over longer distances, however, the pulse undergoes significant broadening and reaches saturation rapidly. 

We explain the nonlinear phase dynamics using a variational approach~\cite{PhysRevA273135,Roy08}, which yields the following evolution equation for the phase of the field (See Supplemental Material for details)
\begin{align} 
\dot{\phi} = (\Omega_0 - \omega_x) - \frac{\beta_2}{\hbar^2 X^2} - \frac{5 A^2 (3 \beta_x + 2 g_x)}{4 \sqrt{2} \hbar }\,.
\end{align} 
The first term represents the LP energy, while the second term accounts for dispersive pulse broadening governed by the group velocity dispersion $\beta_2$ and the time  dependent pulse width $X(t)$. The third term describes the nonlinear contribution, scaling with the squared pulse amplitude $A^2$ and the effective nonlinearities resulting from the average over a Gaussian profile. As illustrated in Fig. \ref{fig3:fihsdsdcsdcsd1mixdscingdfv}(b), the nonlinear term dominates the initial temporal evolution. At later times, GVD-induced broadening becomes significant, eventually leading to phase saturation. In the slow light regime, this saturation is deferred, allowing the nonlinear interaction to drive the phase to substantially larger values.

\emph{Conclusion.} We presented a simple and efficient approach to enhance the nonlinear phase shift in the strong light-matter coupling regime. We consider a corrugated waveguide with defects introduced through variations of the lattice constant, which gives rise to the slow-light regime. Our analysis is based on material parameters relevant to perovskites. In the absence of slow light, achieving an appreciable phase shift is hindered by group-velocity dispersion and the short interaction time, resulting in a very small effect. In contrast, under strong coupling between slow-light photons and excitons, these limitations are alleviated, leading to a nonlinear phase shift enhanced by more than a factor of 20. We expect stronger further improvement when this approach is combined with  material confinement~\cite{arxiv251014566}. It is worth noting that GaAs-based materials may offer somewhat larger phase shifts compared to perovskites; however, as discussed in the Supplementary Material, the difference could be substantial only in the case of a very large superlattice period $N_g$. Importantly, the proposed scheme may operate at room temperature, avoiding the need for the extreme cooling typically required to achieve large nonlinear phase shifts in quantum systems. Note that the moderate nonlinear phase shifts (tens of mrad) reported here can already be useful for applications in quantum computing~\cite{PhysRevApplied.15.054054} and quantum neural networks~\cite{swierczewski2026quantumlight}.

\section*{Acknowledgments}
AR and MM acknowledge funding from the European Union’s Horizon Europe research and innovation programme under grant agreements No. ID 101115575 (Q-ONE) and ID 101130304 (PolArt). The C4QEC project is carried out within the IRAP of the Foundation for Polish Science co-financed by the European Union. 
\section*{DATA AVAILABILITY}
Supporting data used to generate the figures are available at~\footnote{10.5281/zenodo.19350135}.
\bibliography{refs}
\end{document}